\documentclass[12pt,preprint]{aastex}
\usepackage[nonamelimits]{amsmath}

\begin{document}

\title{Spectra of the Expansion Stage of X-Ray Bursts}

\author{Nickolai Shaposhnikov\altaffilmark{1} and Lev
Titarchuk\altaffilmark{2,3}}
\altaffiltext{1}{George Mason University, School for Computational Sciences;
Center for Earth Observing and Space Research, Fairfax, VA 22030; 
nshaposh@scs.gmu.edu}
\altaffiltext{2}{George Mason University/Center for Earth
Observing and Space Research, Fairfax, VA 22030; and US Naval Research
Laboratory, Code 7620, Washington, DC 20375-5352; lev@xip.nrl.navy.mil }
\altaffiltext{3}{NASA/ Goddard Space Flight Center, Greenbelt 
MD 20771, USA; lev@lheapop.gsfc.nasa.gov}

\begin{abstract}
We present an analytical theory of thermonuclear X-ray burst atmosphere 
structure. Newtonian gravity and diffusion approximation are assumed. 
Hydrodynamic and thermodynamic profiles are obtained as a numerical 
solution of the Cauchy problem for the first-order ordinary differential 
equation. We further elaborate a combined approach to the radiative 
transfer problem which yields the spectrum of the expansion stage 
of X-ray bursts in analytical form where Comptonization and free-free 
absorption-emission processes are accounted 
for and $\tau\sim r^{-2}$ opacity dependence is assumed. Relaxation method 
on an energy opacity grid is used to simulate radiative 
diffusion process in order to match analytical form of spectrum, which 
contains free parameter, to energy axis. Numerical and analytical 
results show high similarity. All spectra consist of a power-law soft 
component and diluted black-body hard tail. We derive simple approximation 
formulae usable for mass-radius determination by observational spectra 
fitting. 
\end{abstract}


\section{Introduction}

First discovered by \citet{gri75}, strong X-ray bursts are believed to occur 
due to thermonuclear explosions in the bottom helium-reach layers of the 
atmosphere accumulated by a neutron star during the accretion process in 
close binary system. Since then dozens of burster-type X-ray sources were 
found. One of the distinctive feature of Type I X-ray bursts is the sudden 
and abrupt ($\sim 1$ s) luminosity increase (expansion stage) followed by 
exponential decay (contraction stage). Energy released in X-ray radiation 
during the first seconds greatly exceeds
 the Eddington limit for layers above the helium burning zone 
which are no longer dynamically stable. Super-critically irradiated shells of 
atmosphere start to move outward, producing an expanding wind-like envelope. 
The average lifetime of an X-ray bursts is sufficient for steady-state regime 
of mass loss to be established when the local luminosity throughout the most 
of the atmosphere is equal or slightly greater than the Eddington limit. 

During the last two decades the problem of determining properties of
 radiatively driven winds during X-ray bursts was subjected to extensive 
theoretical and numerical studies. Various  theories were put forward with 
gradually increasing level of accuracy of the problem description, but only a few
approaches addressed the case of considerably expanded photosphere  under influence of 
near-Eddington luminosities \citep{lth,esz,lap,t94}. 
See \citet{lpt} for a detailed review of X-ray burst study during 80's 
and the beginning of 90's. 

Similarly to the problem of accretion flows, notion of the existence of sonic 
point in continuous flow became a natural starting point in the analysis of wind flows 
from stellar objects. \citet{ehs83}, hereafter EHS, investigated the 
structure of the envelopes with steady-state mass outflow and pointed out the 
 higher Eddington luminosity in the inner shells due to the prevalent higher 
temperatures and correspondingly lower Compton scattering opacities.
They showed that the product of opacity and luminosity remains almost constant 
throughout the atmosphere which is the key assumption of the model. 
The existence of wind-like solutions for critically irradiated atmospheres 
was proved. \citet{t94}, hereafter T94, studied analytically spectral 
shapes of the expansion and contraction stages of bursts. 
He showed how EHS's approach to hydrodynamic problem can be greatly simplified 
with the sonic point condition properly calculated and tied with 
conditions at the bottom of the envelope.  \citet{ht} applied 
the T94 model to extract the neutron star mass-radius relations from the 
observed burst spectra in 4U 1820-30 and 4U 1705-44.

\citet{ntl94}, hereafter NTL,  
adopted a high accuracy numerical approach to the problem of X-ray burster atmosphere 
structure based on the moment formalism \citep{trn81,ntz91}. 
They integrated a self-consistent system of frequency-independent, 
relativistic, hydrodynamical and radiative transfer equations over the whole 
atmosphere including the inner dense helium-burning shells. 
Three important characteristic of X-ray burst outflow were obtained 
in this work: the helium-burning zone temperature was maintained 
approximately at the level of  $3 \times 10^9$ K, the 
temperature of the photosphere was shown to depart appreciably 
from the electron temperature and to stay constant 
at the outer shells, and the existence of the maximum and the minimum 
values of the mass loss rate was found.

 One of the goals of these studies was to provide the algorithm 
of determination  of the compact object characteristics by 
analyzing observational data. With
 the advent of high spectral and time resolution  observational instruments 
(such as Chandra, RXTE, USA, XMM-Newton missions) the task of 
obtaining a suitable tool
 for fitting the energy spectra became extremely important. Despite numerous
 earlier studies of X-ray burst observations, recent developments have shown 
a  growing interest of the astrophysical community in this area 
\citep{str,kul}.

 Obviously, the problem of radiative transfer in relativistically moving 
media is very complicated one and under rigorous consideration it must be 
solved numerically. In this paper we develop  an alternative approach which 
allows both numerical and analytical solutions and successfully accounts for 
all crucial physical processes involved. We show how this problem under some 
appropriate approximations yields the spectrum of radiation from spherically 
symmetric outflows in an analytical form. We concentrate on the case of 
extended atmosphere with inverse cubic power dependence of the number density 
on radius, which is more appropriate for the expansion stage but can also be 
employed for description of the contraction as a sequence of models with 
decreasing mass-loss rate.

We represent a numerical approach to the problem which then provides
the validation of our analytical description. We adopt the general 
approach formulated in EHS and developed in T94. The problem of determining 
profiles of thermodynamic variables of steady-state radiatively driven 
outflow was solved in T94. The problem is reduced to the form of a first 
 order differential equation, which allows easy and precise numerical 
solution. For a completeness we present this method in Section 2. Using 
atmospheric profiles obtained for different neutron star configurations we 
solve the problem of radiative transfer by relaxation method on an 
energy-opacity logarithmic grid. We perform temperature profile correction 
 by applying temperature 
equation to the obtained spectral profiles. The basic formulae are given in 
Section 3.1. Then the analytical description of the problem is represented in
 detail. The analytic solution of radiative transfer equation on the 
atmospheric profile $\tau\sim r^{-2}$ is presented in T94. 
Here we review the solution by carrying out the 
integration without introducing any approximations. In Section 4 we compare 
and match our analytical and numerical results to describe the behavior of 
free parameter. We finalize our work by examining the
properties of our analytic solution, combine it with the results
of Section 4 and construct the final formula for fitting the spectra 
in Section 5. The discussion of our method along with some other 
important issues concerning the problem being solved are presented in
Section 6. Conclusions follow in the last Section.

\section{Hydrodynamics}

As we already mentioned, that the  calculation  of X-ray burst spectra can be 
treated as a steady state problem. To justify this assumption one has to 
compare the characteristic times of phenomena considered. Time scale  for the 
photosphere to collapse can be estimated as follows
\[t_{coll}=\int_{r_s}^{r_{ph}}\frac{dr}{v_{coll}},\quad\mathrm{where}\quad v_{coll}\approx 
\sqrt{\frac{2 \,G M_{ns}}{r}(1-l)}.\] 
Here $r_s$ denotes the sonic point radius, which is adopted as the outer 
boundary of the photosphere throughout this paper. Dimensionless luminosity is
 $l=L/L_{Edd}$, where Eddington luminosity is given by
\begin{equation}
\protect\label{hydro1}
L_{Edd}=\frac{4\pi c\,G M_{ns}}{\kappa}.
\end{equation}
Opacity $\kappa$ is expressed by the Compton scattering opacity  with Klein-Nishina correction by \citep{pac}
\begin{equation}
\protect\label{hydro2}
\kappa=\frac{\kappa_{0}}{(1+\alpha T)},
\end{equation}
$\kappa_{0}=0.2(2-Y_{He})~{\rm cm}^{2} g^{-1}$ 
with $Y_{He}$ being the helium abundance, 
and $\alpha=2.2\times10^{-9}K^{-1}$. It is exactly this temperature dependence
 of the opacity that is responsible for the excessive radiation flux, which 
appears to be super-Eddington to the outer less hot layers of the atmosphere.
In the framework of strong X-ray bursts the following condition are usually 
satisfied: $r_s \gtrsim 10^3$km $>>r_{ph}$ , $l\sim 0.99$. Putting 
$m=M_{ns}/M_{Sun}\sim 1$ results in a time of collapse of the order of several 
seconds, the observed time which a Type I X-ray burst usually lasts. For 
evaluation of the time for photons to diffuse through the photosphere, we note
 that a number of scattering events is $N\approx \tau_{ph}^2$ [see, for example, 
 \citet{rbk79}], where $\tau_{ph}$ is the total opacity of the photosphere, 
 which is $\sim 10$. The  time for a photon to escape is
\[t_{esc}\sim\frac{\tau_{ph}^2}{\sigma_T n_e c}\sim 
\frac{r_{ph}}{c}\tau_{ph}\sim 0.1 ~~s.\]
This indicates that the hydrodynamic structure develops at least ten times 
slower than the photons diffusing time through the photosphere. Although 
these time scales can become comparable in cases of greatly extended 
atmospheres, generally steady-state approximation is acceptable.

\subsection{Basic equations for radiatively driven outflow}

The problem of mass loss as a result of radiatively driven wind was 
formulated by EHS. For convenience of the reader we summarize all 
equations important for the derivations in the following sections and 
refer the reader to EHS paper for details. The system of equations describing 
steady-state outflow in spherical symmetry consists of a well known 
Euler (radial momentum conservation) equation 
\begin{equation}
\protect\label{hydro4}
v\frac{d v}{d r}+\frac{G M}{r^{2}}+\frac{1}{\rho}\frac{dP}{dr}=0,
\end{equation}
the mass conservation law
\begin{equation}
\protect\label{hydro5}
\frac{d}{dr}(4\pi r^{2}\rho v)=0,
\end{equation}
the averaged radiation transport equation in the diffusion approximation
\begin{equation}
\protect\label{hydro6}
\kappa L_{r}=-\frac{16\pi a c r^{2} T^{3}}{3\rho}\frac{dT}{dr},
\end{equation}
and the entropy equation
\begin{equation}
\protect\label{hydro7}
vT\frac{ds}{dr}+\frac{1}{4\pi r^{2}\rho}\frac{dL_{r}}{dr}=0,
\end{equation}
where $P,\rho,T,s$ and $L_{r}$ are, respectively, the pressure, 
the density, the temperature, the specific entropy, and the diffusive energy 
flux flowing through a shell at $r$.

The outflowing gas is taken to be ideal.
Dimensionless coordinate $y$, which is the ratio of the radiation pressure $P_{r}$, to the gas pressure $P_{g}$, is introduced by
\begin{equation}
y=\frac{P_{r}}{P_{g}}=\frac{\mu m_p}{k}\frac{aT^{3}/3}{\rho},
\end{equation}
where $\mu=4/(8-5 Y_{He})$ and $m_p$ are the mean molecular weight and the
 mass of proton, respectively. Then the other thermodynamic quantities are 
expressed in terms of $y$ and $T$ as
\begin{equation}
\protect\label{hydro9}
P=P_{r}+P_{g}=\left(1+\frac{1}{y}\right)\frac{aT^{4}}{3},
\end{equation}

\begin{equation}
\protect\label{hydro10}
\rho=\frac{a\mu m_p}{3 k}\frac{T^3}{y},
\end{equation}

\begin{equation}
\protect\label{hydro11}
s=\left(\frac{k}{\mu m_p}\right)[4y+\ln y-(3/2)\ln T],
\end{equation}

\begin{equation}
\protect\label{hydro12}
h=\frac{k}{\mu m_p}(4y+5/2)T,
\end{equation}
where $h$ is the specific enthalpy. 

Integrals of equation (\ref{hydro5}) and (\ref{hydro4}) give energy flow 
and mass flux correspondingly
\begin{equation}
\protect\label{hydro13}
4\pi r^{2}\rho v=\Phi,
\end{equation}
\begin{equation}
\protect\label{hydro14}
\left(\frac{ v^{2}}{2}-\frac{G M}{r}+h\right)\Phi+L_{r}=\Psi.
\end{equation}

To make two more integrations, which can not be performed analytically, the 
constancy of $\kappa L_{r}$, which stands for the integral of equation 
(\ref{hydro6}), over the relevant layers is assumed. In EHS this assumption 
is confirmed numerically. We can also justified by the following 
consideration. At the near-Eddington regime the radiation pressure
$aT^4/3$ is much greater than the pressure of gas almost everywhere except 
for the innermost layers adjacent to the helium-burning zone. Neglecting the
 gas pressure in equation (\ref{hydro4}) and multiplying it by 
$-r^2$ we get
\begin{equation}
\protect\label{hydro14_1}
\frac{\kappa  L_{r}}{4\pi c}=GM+r^2v\frac{dv}{dr}.
\end{equation}
Here we moved first two terms of (\ref{hydro4}) to the right hand side and 
used equation (\ref{hydro6}) to express third term by $\kappa  L_{r}$. For 
inner and intermediate layers of the atmosphere the last term in 
(\ref{hydro14_1}) is negligible and equation reduces to $\kappa L_r=\kappa_0 L_{0}$. 
This term can become considerably large for the outermost layers where $L_r$ must 
exceed $L_{Edd}$. This also is in agreement with observations of X-ray bursts 
from which super-Eddington luminosities are inferred. For the sake of 
analytical consideration we consider $\kappa L_{r}$ to be constant 
throughout the whole atmosphere, and the third integral is 
\begin{equation}
\protect\label{hydro15}
\kappa L_{r}=\kappa_{0} L_{0}=const.
\end{equation}
Replacing $L_{r}$ of equation (\ref{hydro7}) with equation 
(\ref{hydro15}), the fourth integral is obtained as
\begin{equation}
\protect\label{hydro16}
\Phi s + \alpha L_{0}\ln T=\Xi=const.
\end{equation}

Boundary conditions need to be imposed at the bottom and the outer boundary to 
determine the four integration constants $\Phi,\Psi,L_{0}$, $\Xi$ and to obtain
 a specific solution.

At the bottom of the atmosphere close to the helium-burning zone there should 
be a point where gas and radiation pressure are equal. As another important
numerical result EHS showed that near the neutron star surface the 
temperature and radius profiles level off with respect to $y$, so there is 
always a point where
\begin{equation}
\protect\label{inbcond}
r=r_{b},~ T=T_{b}, ~y=1,
\end{equation}
and $r_{b}$ is well approximated by the radius of the neutron star $R_{ns}$. 
However, $T_b$ can not be considered as a real temperature of helium-burning 
shell at the bottom of the star surface  because thermonuclear processes 
are not included in the model. Rigorous account of helium-burning 
NTL shows that temperature of burning shells vary 
in small range of values. 

To obtain the outer boundary condition the concept of sonic point is used. 
For the solution to be steady-state and to have finite terminal 
velocity it should pass sonic point where 
\begin{equation}
\protect\label{hydro21}
v_{s}^{2}=\frac{G M_{ns}}{2r_{s}}=
\left(\frac{\partial P_{s}}{\partial\rho_{s}}\right)_{\Xi}=
\left(\frac{k}{\mu m_p}\right)Y_{s}T_{s},
\end{equation}
\begin{equation}
\protect\label{hydro22}
Y_{s}=\frac{\lambda+4(1+y_s)(1+4y_s)}{\lambda+3(1+4y_s)},
\end{equation}
where $\lambda$ is a quantity related to the ratio of  the energy flux to 
the mass flux (see formula 22, below). 
In EHS this formula contains a typo. We give a proper derivation of this 
form for $Y_s$ in Appendix C.

\subsection{ODE solution of the hydrodynamic problem}

T94 has shown how the treatment of the hydrodynamic problem can be reduced 
to a Cauchy problem with the boundary condition determined at the sonic point. 
This treatment provides a  high-accuracy method of obtaining the 
hydrodynamic solution. The crucial point is to relate the position of the 
sonic point with the values of the velocity and the thermodynamic quantities 
before solving the set of appropriate hydrodynamic equations. The profile 
of the expanded envelope is then obtained as a result of the integration of a 
single  first-order ordinary differential equation (ODE) from the sonic point 
inward up to the neutron star surface. For completeness we present 
the details of this approach.

At the bottom of the atmosphere the potential energy per unit mass of the gas, $GM/r$, is 
significantly greater than the kinetic energy, $v^2/2$, and enthalpy. 
Therefore, by ignoring these terms in equation (\ref{hydro14}), we obtain the value of the mass flux
\begin{equation}
\Phi=\frac{R_{ns}}{G M_{ns}}\alpha T_{b}L^{0}_{Edd}.
\end{equation}
The inner boundary condition (\ref{inbcond}), the integral (\ref{hydro16}), 
and  equation (\ref{hydro11}) for entropy can be used to find the temperature distribution with respect to $y$
\begin{equation}
\protect\label{hydro24}
T=T_{b}\,y^{-1/\lambda}\exp \left[-\frac{4(y-1)}{\lambda}\right],
\end{equation}
and 
\begin{equation}
\lambda=\frac{\alpha\mu m_p}{k\Phi}L_{0}-\frac{3}{2}.
\end{equation}
The condition at the sonic point (\ref{hydro21}) allows us to find the constant $\Psi$
\begin{equation}
\Psi=h_{s}\Phi+L_{r}(r_{s})-\frac{3}{4}\frac{G M}{r_{s}}\Phi.
\end{equation}
Combining the mass and energy conservation laws (\ref{hydro13}) and (\ref{hydro14}), the specific enthalpy and density equations (\ref{hydro10}) and (\ref{hydro12}) and 
 eliminating the radial coordinate $r$ between equations (\ref{hydro13}) and (\ref{hydro14}) yield the following dependence of the velocity derivative $v'$ with respect to $y$
\begin{equation}
\protect\label{hydro26}
v'_y(y,v)=v\left[\left(1+3\frac{1+4y}{\lambda}\right)\frac{1}{y}-75.2\frac{rT(8-5Y_{He})(1+4y)(1+\alpha T)}{\lambda r_{b,6}T_{b,9} (\Psi/\Phi-v^2/2-h+G M_{ns}/r)}\right].
\end{equation}
Derivation of equation (\ref{hydro26}) is given in Appendix D.

By imposing boundary conditions at the bottom of the extended envelope 
(at the neutron star surface) and at the sonic point, we can determine the four 
integration constants necessary to obtain a specific solution. 
One can note the obvious fact that the bottom of the envelope can not serve 
as a starting point of integration of equation (\ref{hydro26}) as long as
 $v_{b}=0$, which introduces uncertainty. Fortunately, we can calculate 
parameters at the sonic point in the framework of our problem 
description by solving a nonlinear algebraic equation, which involves 
only $y_s$, the ratio of the radiation pressure $P_{r}$ to the gas 
pressure $P_{g}$ at that point. Substitution of the radial coordinate 
$r_{s}$ and velocity $v_s$ from the definition of the sonic point 
position (\ref{hydro21}), and the sonic point density, $\rho_{s}$ from 
equation (\ref{hydro10}), we find
\[r_s=\frac{G M_{ns}\, \mu m_p}{2 k Y_s T_s},\hspace{0.1in}v_s=
\left(\frac{k}{\mu m_p}Y_s T_s\right)^\frac{1}{2},\hspace{0.1in}
\rho_s=\frac{a\mu m_p}{3 k}\frac{T_s^3}{y_s},\]
and the expression for the temperature given in equation (\ref{hydro24}) 
into the mass conservation law (\ref{hydro13}), after some algebra give an 
equation for the value of $y_{s}$
\begin{equation}
\protect\label{hydro27}
y_{s}=\frac{\lambda}{4}\ln \left\{\left[\frac{(2-Y_{He})m^2}{r_{b,6}T_{b,9}}\right]^{2/3}\frac{T_{b}}{0.149\,(8-5Y_{He})^{5/3}Y_s y_s^{1/\lambda+2/3}}\right\}+1.
\end{equation}
Here $r_{b,6}$ and $T_{b,9}$ are the neutron star surface radius and temperature in units of $10^6$ cm and $10^9$ Kelvin respectively. Since $Y_s$ is expressed in terms of $y_s$ (eq. \ref{hydro22}), 
equation (\ref{hydro27}) can be solved to determine the value of $y_s$. 
Knowledge of $y_s$ can then, by substitution in  equation (\ref{hydro24}), 
yield the value of the temperature at the sonic point $T_s$ and then $v_s$ 
from  equation (\ref{hydro21}). It is now possible to relate $v_s$ to 
$T_s$, $T_b$, $r_s$, $r_b$, thus obtaining the analytical expression 
for the various dynamical quantities at the sonic point in terms 
of the values of the parameters associated with the boundary conditions.
To obtain the solution of the hydrodynamical problem for a particular set of 
input parameters we use a standard Matlab/Octave package function minimizators
 and ODE solvers.  

\section{Radiative Transfer Problem}

The radiation field of X-ray burst atmosphere may be described by the 
diffusion equation, written in spherical geometry, with the Kompaneets's 
energy operator (see T94):

\[
\hspace{-1in}\frac{1}{3} \left(\frac{\partial^2 J_\nu}{\partial \tau^2}-
\frac{2}{r \alpha_T}\frac{\partial J_\nu}{\partial \tau} \right)=
\frac{\alpha_{ff}}{\alpha_T}(J_\nu-B_\nu)-\]

\begin{equation}
\protect\label{rad1}
\hspace{0.4in}-\frac{k T_e}{m_e c^2}\,x_0\frac{\partial}{\partial x_0}
\left(x_0\frac{\partial J_\nu}{\partial x_0}-3J_\nu+\frac{T_0}{T}J_\nu\right),
\end{equation}
where $x_0=h\nu/kT_0$ is a dimensionless frequency, $T_0$ being the effective temperature; 
$\alpha_{ff}$ and $\alpha_T=\sigma_T n_e$ are the coefficients of free-free absorption and 
Thompson scattering, respectively, whose ratio is given by \citep{rbk79}
\begin{equation}
\protect\label{affat}
\frac{\alpha_{ff}}{\alpha_T}=1.23\, \rho g_{14}^{7/8}(1-Y_{He}/2)^{7/8} \Psi(x_0) \left(\frac{T_0}{T}\right)^{1/2}\end{equation}
with
\[\Psi(x_0)=\frac{\tilde{g}(x_0 T_0/T)}{x_0^3}\left(1-e^{-x_0T_0/T}\right).\]
Here $\tilde{g}(x)$ is the Gaunt factor \citep{gr}
\[\tilde{g}(x)=\frac{\sqrt{3}}{\pi}\,e^{x/2}K_0(x/2),\]
$K_0(x)$ is the Macdonald function, and $g_{14}$ denotes the 
free-fall acceleration onto the neutron star surface, in units of $10^{14}$ cm s$^{-1}$. 

 We combine equation (\ref{rad1}) with the outer boundary condition of zero energy inflow
\begin{equation}
\protect\label{rad3}
\left(\frac{\partial J_\nu}{\partial \tau}-\frac{3}{2}J_\nu\right)\Bigl|_{\tau=0}=0
\end{equation}
and the condition of equilibrium blackbody spectrum at the bottom of the photosphere, which is represented in 
a dimensionless form as
\begin{equation}
B_\nu=\frac{x_0^3}{\exp(x_0T_0/T)-1}.
\end{equation}

We will make use of the temperature equation, which is obtained by integration of (\ref{rad1}) over frequency. 
The opacity operator vanishes as a result of the total flux conservation with respect 
to optical depth, leaving us with
\[\hspace{-1.5in}\frac{kT_e}{m c^2}\left(4\int_0^\infty J_\nu dx_0 - 
\frac{T_0}{T}\int_0^\infty x_0 J_\nu dx_0\right)=\]
\begin{equation}
\protect\label{rad4}
\hspace{0.5in}=1.23\, \rho g_{14}^{7/8}(1-Y_{He}/2)^{7/8}
\left(\frac{T_0}{T}\right)^{1/2}\left[\int_0^\infty J_\nu 
\Psi(x_0) dx_0-\frac{2\sqrt{3}}{\pi}\frac{T}{T_0}\right].
\end{equation}
In the condition of the extended photosphere of X-ray bursts, the density 
usually is very low and the left-hand side of the last equation can be 
neglected reducing the last equation to the formula for temperature 
\begin{equation}
\protect\label{rad5}
\frac{T}{T_0}=\frac{1}{4}\left(\int_0^\infty x_0 J_\nu dx_0\Bigl/\int_0^\infty J_\nu dx_0\right),
\end{equation}
We will use the last relation to produce a corrected temperature profile for 
the photosphere where it departs significantly from that given by the 
hydrodynamic solution. 

\subsection{Analytic Description of Radiative Diffusion}

Hydrodynamic profiles calculated in section 2 show that during the expansion
 stage in the vicinity of the sonic point $v\sim v_s(r/r_s)$. Considering this
 relation to be true throughout the entire envelope, according 
to the mass conservation law, we can write 
\begin{equation}
\protect\label{rad6}
n_e=\frac{\rho}{m_p}\left(1-\frac{Y_{He}}{2}\right)=\frac{\Phi}{4\pi m_p v r^2}=\frac{GM\Phi}{8\pi v_s^3m_p}\left(1-\frac{Y_{He}}{2}\right)r^{-3},
\end{equation}
where $\Phi$ is the mass loss rate and $v_s$ is the  velocity of gas at the 
sonic point. In this case opacity can be expressed as
\begin{equation}
\tau=C \int_r^\infty \frac{\sigma_T}{r^3}dr=\frac{C\sigma_T}{2r^2}.
\end{equation}
Noting that in this case 
\begin{equation}
\protect\label{rad7}
\tau=\frac{r\alpha_T}{2},
\end{equation}
 we can rewrite the radiation transfer equation in the form
\begin{equation}
\protect\label{rad8}
\frac{\partial}{\partial \tau}\frac{1}{\tau}\frac{\partial J_\nu}{\partial \tau}=
\frac{3}{\tau}\frac{\alpha_{ff}}{\alpha_T}(J_\nu-B_\nu)-\frac{3kT_e}{m_e c^2 \tau}L_\nu(J_\nu).
\end{equation}

The boundary conditions are given by
\begin{equation}
\protect\label{rad9}
J_\nu|_{\tau=\tau_{th}}=B_\nu(\tau_{th})
\end{equation}
at the inner boundary of photosphere, and 
\begin{equation}
\protect\label{rad10}
H=\frac{4\pi}{3}\int_0^\infty\frac{\partial J_\nu}{\partial \tau}d\nu \Bigl|_{\tau=0}=\frac{L}{4\pi R_s^2}
\end{equation}
at the sonic surface. The ratio $\alpha_{ff}/\alpha_T$ can be written in the form
\[\hspace{-0.5in}\frac{\alpha_{ff}}{\alpha_T}=1.23\left(1-\frac{Y_{He}}{2}\right)^{-5/8}\left(\frac{2m_p}{\sigma_T}\right)^{3/2}\left(\frac{8\pi v_s^3}{GM\Phi}\right)^{1/2}\]
\begin{equation}
\protect\label{rad11}
\times g_{14}^{-7/8}\frac{\tilde{g}(x)(1-e^{-x})}{x^3}
\left(\frac{T_0}{T}\right)^{7/2}\tau^{3/2}=D\Psi(x)\tau^{3/2},
\end{equation}
where $x=h\nu/kT_e$, and $\Psi(x)=\tilde{g}(x)(1-e^{-x})/x^3$.

Stated in this way the problem of radiative transfer allows an analytical approach. The solution of the radiative transfer equation (\ref{rad8}) is
\begin{equation}
\protect\label{rad12}
J(t,x)= B_\nu \frac{t^{8/7}}{2^{4/7}
\Gamma\left(\frac{11}{7}\right)}
\left[\frac{\Gamma\left(\frac{3}{7}\right)}
{2^{4/7}}+\frac{8}{7}\,t_{th}^{-4/7}K_{4/7}(t_{th})\right],
\end{equation}
where $K_p$ is the modified Bessel function of the first kind, and
\begin{equation}
\protect\label{rad13}
t=\frac{4}{7}\sqrt{3D\Psi(x)}\,\tau^{7/4}.
\end{equation}
Details of derivation of this formula are given in Appendix A.

\subsection{Evaluation  of $\tau_{th}$ and $T_c$}

The next step is to find the color temperature and to determine the 
thermalization depth $\tau_{th}$ where the boundary condition (\ref{rad9}) 
is valid. For saturated Comptonization, the occupation number behaves 
in accordance with Bose-Einstein photon distribution $n=(e^{\mu+x}-1)^{-1}$ 
which might be described as a diluted blackbody spectrum or as a diluted Wien 
distribution.

At first we evaluate the color temperature assuming a blackbody spectral 
shape. We look for the solution of the form
\begin{equation}
\protect\label{rad14}
n(\tau,x)=\frac{R(\tau)}{e^x-1}.
\end{equation}
The solution, which is described in detail in Appendix B, gives for $R(\tau)$
\begin{equation}
R(\tau)=1-\frac{2^{3/7}}{\Gamma\left(\frac{4}{7}\right)}\frac{\tau}{\tau_{th}} K_{4/7}\left(\left[\frac{\tau}{\tau_{th}}\right]^{7/4}\right).
\end{equation}

As long as $R(\tau)=1$ for $\tau>\tau_{th}$, there is radiation equilibrium 
for optical depths deeper than the photospheric envelope. The temperature 
equation in the zone $0<\tau<\tau_{th}$ reads
\[
\left(\frac{T}{T_0}\right)^4=\frac{2H_0}{R^2}\left(\frac{3}{2\tau_R}\int_0^\tau\tau d\tau+2\right)\Bigl/\frac{\pi^4}{15}R(\tau)
\]
where $H_0=(4\pi R_{ns}^2\pi^5/15)/16\pi^2$ and $\tau_R$ ($<1$) is the optical depth coordinate
at the outer boundary of the expanded atmosphere, $r=R$ (see eq. 35).
This equation can be rewritten as follows 
\begin{equation}
\left(\frac{T}{T_0}\right)^4=\frac{3\tau^2/4+2\tau_R}{2\tau_{ns}R(\tau)}.
\end{equation}
Neglecting $\tau_R$ with respect to $\tau$ and making use of Taylor expansion (\ref{B6}) 
of $R(\tau)$ we get a constant value of the temperature
\begin{equation}
\protect\label{rad15}
\left(\frac{T}{T_0}\right)^4=\frac{3}{8}\frac{2^{8/7}\Gamma\left(\frac{11}{7}\right)}{\Gamma\left(\frac{3}{7}\right)}\frac{\tau_{th}^2}{\tau_{ns}}=0.356\frac{\tau_{th}^2}{\tau_{ns}}.
\end{equation}
Using the notation
\begin{equation}
\protect\label{rad16}
p=2\tilde{g}(x_*)\approx \ln\left(\frac{2.35}{x_*}\right)
\end{equation}
formula (\ref{rad11}) becomes
\begin{equation}
\protect\label{rad16_1}
D=D_0\left(\frac{T_0}{T}\right)^{7/2}
\end{equation}
where
\[
D_0=1.23\left(1-\frac{Y_{He}}{2}\right)^{-5/8}\left(\frac{2m_p}{\sigma_T}\right)^{3/2}\left(\frac{8\pi v_s^3}{GM\Phi}\right)^{1/2}g_{14}^{-7/8},
\]
while we can write for $\tau_{th}$ 
\begin{equation}
\protect\label{rad17}
\tau_{th}=\left[\frac{4}{7}\sqrt{\tilde{D}}\right]^{-4/7}=\left[\frac{6}{49}p^2D\right]^{-2/7}=\frac{T}{T_0}\left[\frac{6}{49}p^2\right]^{-2/7}D_0^{-2/7}.
\end{equation}
Substituting it into (\ref{rad15}) we get for $T/T_0$
\begin{equation}
\frac{T}{T_0}=0.596\left[\frac{6}{49}p^2\right]^{-2/7}
\frac{D_0^{-2/7}}{\tau_{ns}^{1/2}}.
\end{equation}
Assuming the same electron number density as in (\ref{rad6}) we than express 
opacity at the neutron star surface  $\tau_{ns}$ in the form
\begin{equation}
\protect\label{rad18}
\tau_{ns}=\left(1-\frac{Y_{He}}{2}\right)
\left(\frac{\sigma_T}{2m_p}\right)\left(\frac{GM\Phi}
{8\pi v_s^3 R_{ns}^2}\right).
\end{equation}
If we use the dependence of input parameters $g_{14}$ and $\Phi$ on 
$m,r_{b,6}, T_{b,9}$ and $Y_{He}$, the next useful equations for the color ratio $T/T_0$, 
color temperature $kT$, and thermalization depth $\tau_{th}$ are found:
\begin{equation}
\protect\label{rad20}
\frac{T}{T_0}=\frac{0.191(2-Y_{He})^{1/28}\,r_{b,6}^{1/7}v_{s,8}^{15/14}}{m^{3/28}\,T_{b,9}^{5/14}p^{4/7}},
\end{equation}
\begin{equation}
\protect\label{rad21}
kT=0.4\,m^{1/7}r_{b,6}^{-5/14}\,T_{b,9}^{-5/14}v_{s,8}^{15/14}
(2-Y_{He})^{-3/14}p^{-4/7}\quad\mathrm{keV}\quad,
\end{equation}
\begin{equation}
\protect\label{rad22}
\tau_{th}=90.5\,m^{2/7}r_{b,6}^{-3/14}\,T_{b,9}^{-3/14}v_{s,8}^{9/14}(2-Y_{He})^{1/14}p^{-8/7}.
\end{equation}
Here $v_{s,8}$ is the sonic point velocity in units of $10^8$ cm/s. These relations present the final results of our analytical approach. 
There is still a lack of completeness due to the presence of $p$ and $v_s$ 
in the left-hand sides of this system of equations. Parameter $p$ and sonic
 point velocity are not independent parameters of the problem, but at this 
point they can not be inferred from further analytical consideration. 
Fortunately, these quantities can be quite well approximated by a power
 dependence from $m,r_{b,6},T_{b,9}$ and $Y_{He}$, which is done in the next chapter.

\section{Numerical Results and Comparison with Analytical Description of Radiative Transfer Problem}

\begin{figure}
\plottwo{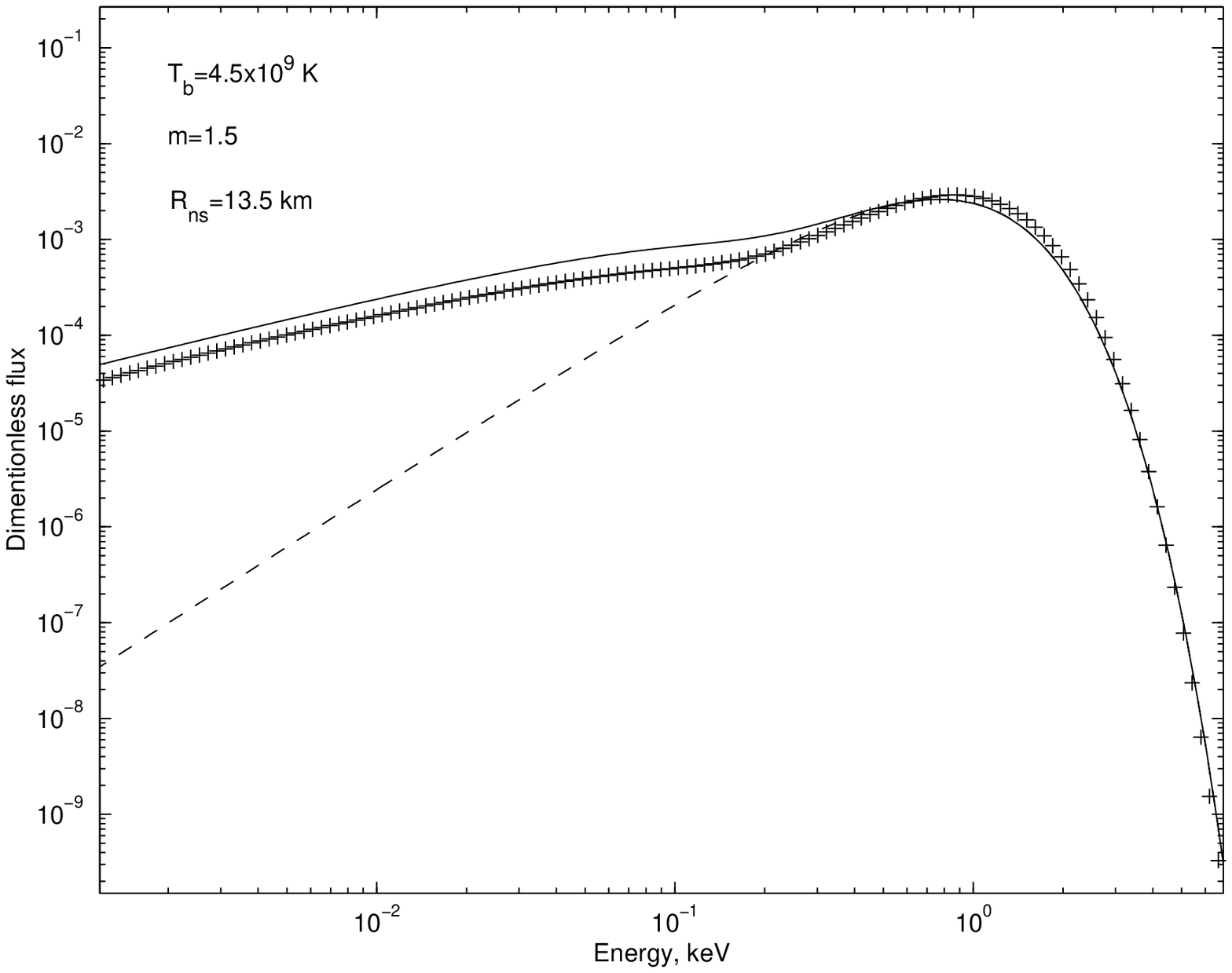}{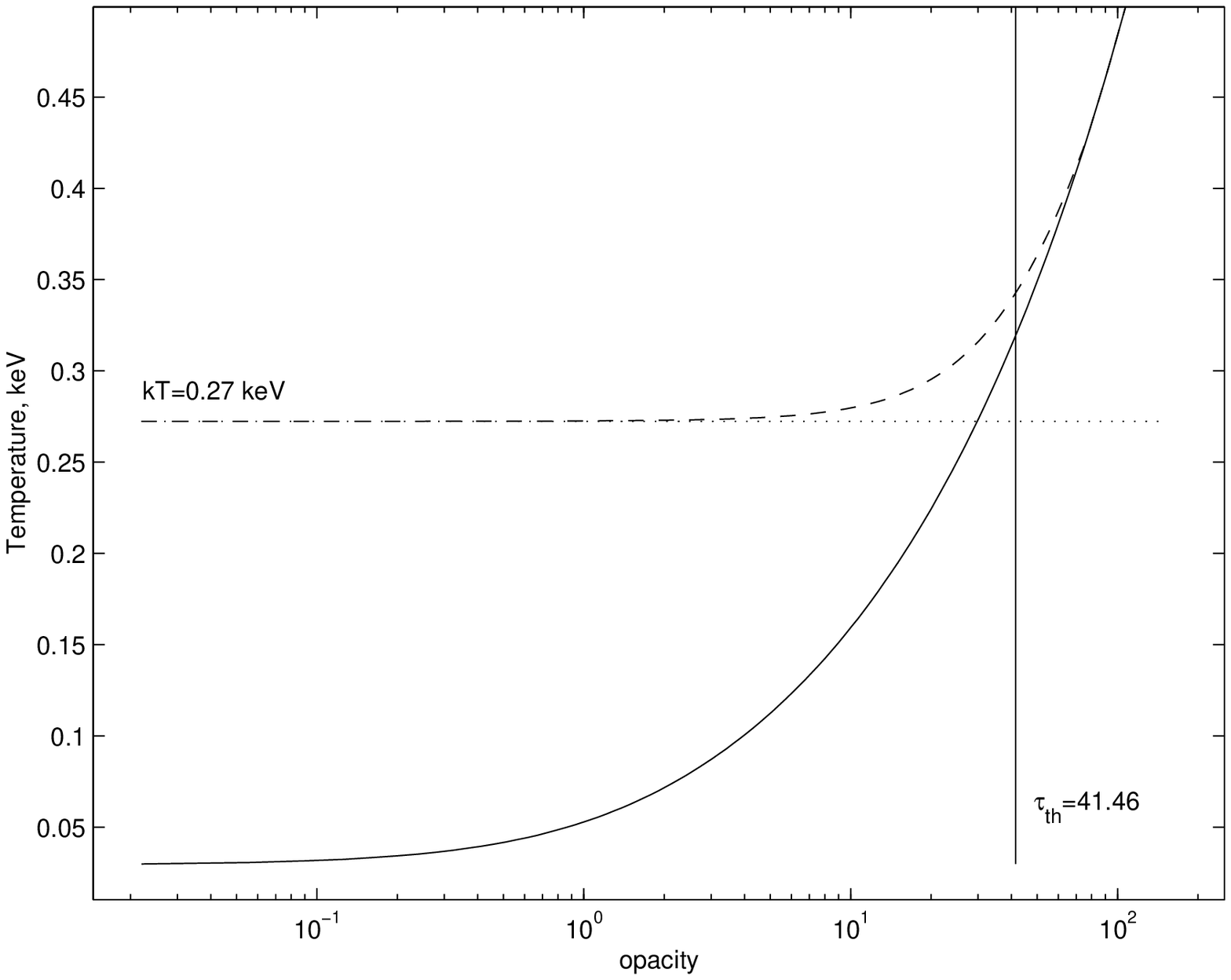}
\caption{Examples of spectra (left) and temperature profiles (right) obtained 
for model with $m=1.5$, \mbox{$R_{ns}=13.5$ km}, $Y_{He}=1$. 
On the left  solid line  represents the analytical solution, dashed 
line indicates the diluted blackbody level, '+' - results of relaxation method simulation. 
On the right  solid line 
is the temperature profile obtained from the initial hydrodynamical solution, 
dashed line is the corrected profile (see text), 
dotted line is analytically calculated color temperature level.\protect\label{fig1}}
\end{figure}

\begin{figure}
\plottwo{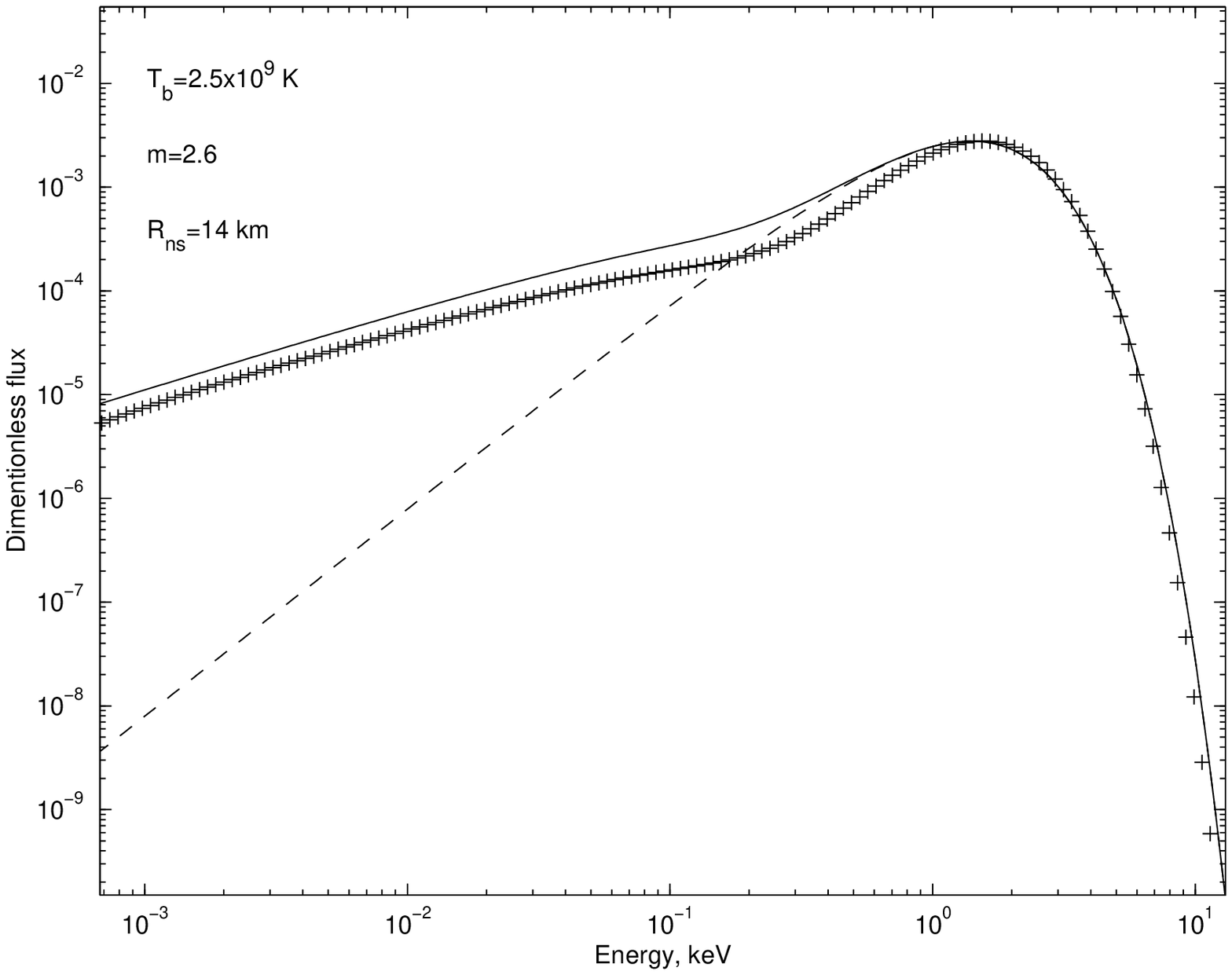}{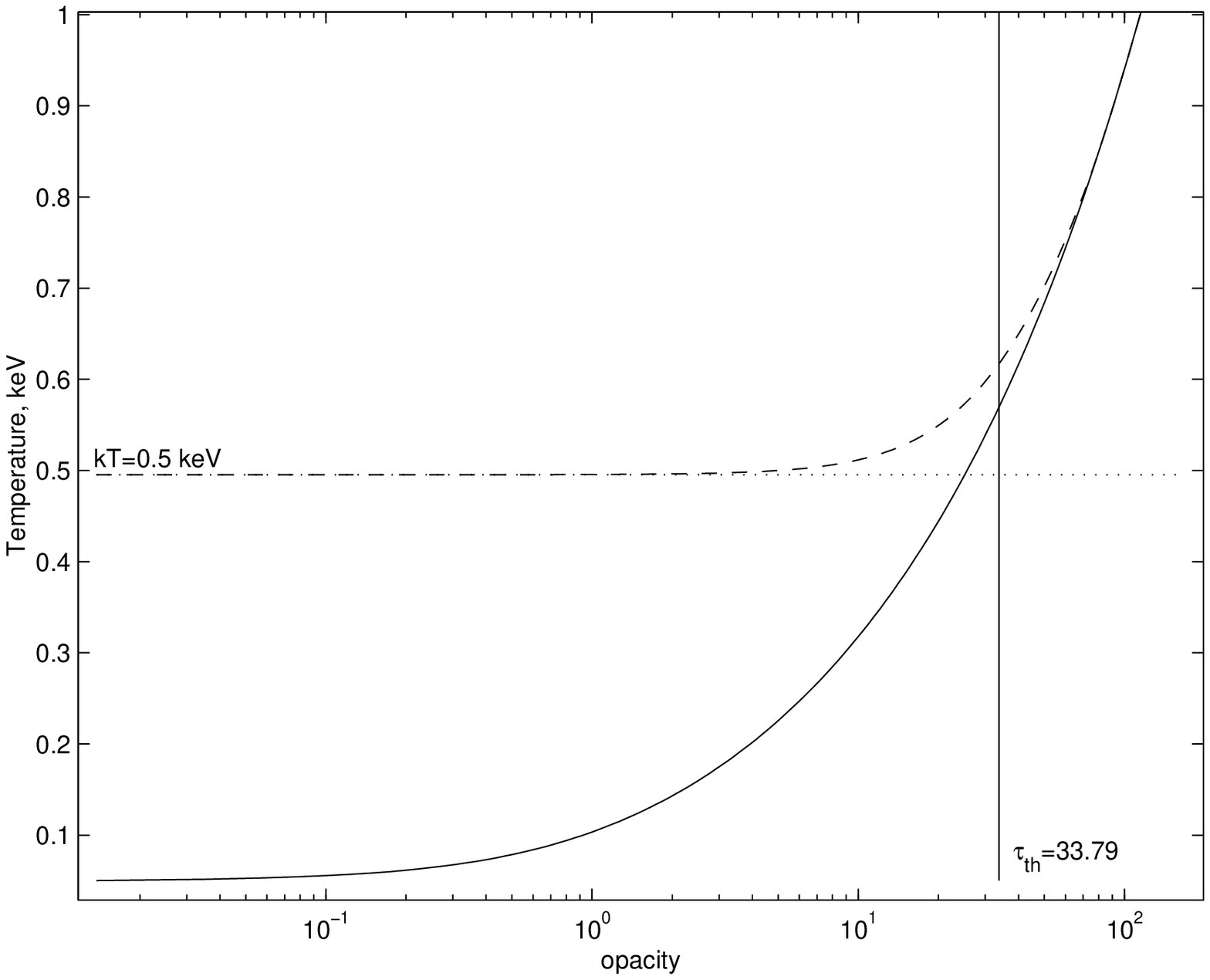}
\caption{Same as Fig.\ref{fig1} but for the model with $m=2.6$, $R_{ns}=14$ km, $Y_{He}=1$.}
\end{figure}

To confirm the validity of our analytical approach and to examine the 
behavior of $p$ and $v_s$ in dependence of different input parameters of 
the problem, we perform numerical modeling of the steady-state radiative 
transfer process. The whole procedure consisted of three steps. 

First, for particular model of neutron star, i. e. for a given mass and 
radius, we  obtain a set of model atmospheres for a chosen set of bottom 
temperatures. These 
solutions provide us with runs of thermodynamical and hydrodynamical profiles,
 sonic point characteristics, masses of the extended envelopes and their loss 
rates. Second, we solve  radiative  transfer equation (\ref{rad1}) on each model atmosphere obtained
 with the relaxation method  (e.g. Press et al. 1992) on an energy-opacity grid 
using logarithmic scale on both dimensions. The energy range included 500 grid 
points. The number of grid points in opacity varied between 100 and 300. 
The opacity domain included the range $\tau_s<\tau<\tau_{max}$, where $\tau_s$ 
is opacity at the sonic point and $\tau_{max}$ was taken large enough to meet safely 
inequality $\tau_{max}>\tau_{th}$. We used the mixed outer boundary condition 
(\ref{rad3}). Spectrum at the inner bottom  of the photosphere was taken a pure black body $B_{\nu}$. 
Numerical calculations of frequency-dependent radiation field consisted of 
two runs of our relaxation code. The first run was performed on 
temperature continuum, obtained from the hydrodynamical solution (see 
section 2.5). Then we calculated a spectral temperature profile using 
formulae (\ref{rad5}) which exhibits a quite expected behavior. At some 
region this corrected profile departs from initial temperature profile 
and levels off at some constant value in absolute agreement with analytic 
result of section 3.2. It is also in a qualitative agreement with the NTL 
self-consisted calculation of radiation driven wind  structure of X-ray 
 burster. A second run is performed on the corrected profile to get more 
reliable spectrum shape. At the final third step we compared analytical 
and numerical solutions. The sonic point provided a natural position  to 
match numerical and analytical solutions. Combining the sonic point parameters,
 calculated through the solution of equation (\ref{hydro27}) and using 
relation (\ref{rad7}) we get for the opacity at the sonic point
\begin{equation}
\protect\label{num1}
\tau_s=\frac{\sigma_T}{2m_p}r_s\rho_s\left(1-\frac{Y_{He}}{2}\right).
\end{equation} 

We calculated and and plotted fluxes given by both methods at the sonic 
point. A particular value of parameter $p$ for analytic model was obtained 
by matching value of $kT$ and corrected level of numerically achieved value
 of photospheric temperature. 

We obtained results for approximately 150 different sets of values 
$T_b, R_{ns}, M_{ns}$ and $Y_{He}$. 
Examples of numerical calculations of spectra for different neutron star 
models  and fitting them with analytical shapes are presented in Figures 1
 and 2. Analytical and numerical shapes match quite well in the wide 
range of neutron star surface temperatures and both show two distinctive 
features of outgoing spectrum of expansion stage: diluted 
black-body like high-frequency component and power law soft excess at the 
lower part. Dependence of the sonic point opacity presented by (\ref{num1}) 
describes correctly the dilution process indicating the assumption of atmosphere 
structure adopted at the analytical model is correct.

Tables 1 and 2,  which summarize results for two different neutron star 
models, are given in order to compare our results with more rigorous  
calculations (NTL). Taking mass-loss rate as an input parameter, 
NTL obtained profiles of different quantities throughout the whole atmosphere.
 They argue that the temperature of burning shell is maintained around 
$3 \times 10^9$ K for all models. 
The temperature of photons departs appreciably from the temperature of
 ambient matter above photospheric radius and stays practically constant
indicating  that radiation becomes essentially decoupled from expanded media. 
We changed the bottom temperature in a wide range of values and inferred
the mass-loss rate, the  mass of envelope, etc.

 Our results are in qualitative agreement with NTL. The crucial
 physical parameters which define the main spectral signatures are the photospheric
radius $r_{ph}$ and its temperature $kT$. Runs of the atmospheric profiles obtained by
both approaches are quite 
similar, although $T_{ph}$ in NTL results is usually $15\%\sim25\%$ greater
than in our models. This difference is explainable. We match isothermal
levels given by numerical and analytical calculations and define the obtained
value as a photospheric temperature. This is the lowest estimation, because 
the temperature profile starts to grow before the bottom of the photosphere
is reached. NTL define $T_{ph}$ as a matter temperature at $r_{ph}$.
A temperature level calculated at the thermalization depth $\tau_{th}$
should compensate the considered difference. The difference in
density profiles, which can achieve a factor of two, will affect
 the spectrum only in the soft part ($\leq 0.2$ keV)
 where the normalization of the power law component can be changed. This fact
does not diminish the validity of our results. The soft component of the spectrum 
can be represented as an independent fitting shape with a normalization 
included as an additional fitting parameter. This matter is not crucial 
at the moment due to the restricted spectral capabilities of current 
X-ray observing facilities.
 One can also notice a quick decrease of the envelope mass, and 
 point out a wide variation of $T_b$. 
 This discrepancy can be explained by differences in model formulations. 
 Specifically, NTL included helium-burning shells into the model and 
 put the inner boundary condition on the ``real'' neutron star surface 
 while our model stops where radiation and gas pressures are equal ($y=1$), 
 which is close but still outside of the helium-burning shell. In our approach, 
 part of the bottom of the atmosphere is left out.  In fact, the lower
the mass-loss rate, the greater the portion of mass missing beyond the point where $y=1$.
 This is clearly seen from the tables.
 The  temperature at the bottom can be considered as an ``effective'' instead
  of the real temperature of  helium-burning zone.

As we have already point out that one needs to know the dependencies 
of $v_s$ and $p$ on input parameters to complete the analytical description and 
thus to employ these results to the fitting of observational X-ray spectra.
Analysis of $v_s$ and $p$ runs show that $\log v_s$ and $\log p$ are 
linear functions of $\log T_b$, $\log R_{ns}$, $\log M_{ns}$ and 
$\log(2-Y_{He})$. We combine all experiments and fit
$v_{s,8}$ and $p$ with a model 
$cons\times T_{b,9}^\alpha r_{b,6}^\beta m^\gamma (2-Y_{He})^{\eta}$ 
by the least squares method  to get
\begin{equation}
\protect\label{num2}
p=7.69\, T_{b,9}^{-0.84} r_{b,6}^{-0.89} m^{0.69} (2-Y_{He})^{-0.22}
\end{equation}  
\begin{equation}
\protect\label{num3}
v_{s,8}=5.46\, T_{b,9}^{-0.71} r_{b,6}^{-0.87} m^{0.63} (2-Y_{He})^{-0.22}
\end{equation}  
with maximum errors of parameters less than 1\%. The ranges of parameters
 included in fitting are 0.3-7.0 for $T_{b,9}$, 0.6-2.0 for $r_{b,6}$, 0.8-2.7
 for $m$ and 0.3-1.0 for $Y_{He}$. These results can be used to substitute
 $p$ and $v_{s,8}$ in equations (\ref{rad20})-(\ref{rad22}). 
Now we have consistent system of equations, which should yield X-ray spectrum 
of burster in the form  of function of only input physical parameters, 
i. e. neutron star mass, radius, surface temperature and elemental abundance. 
 
\section{ Final form of the profile for spectral fitting}

The fact that spectra obtained are a black-body like almost everywhere except 
for small values of energies allows us to proceed with simplification of the 
formula (\ref{rad12}). First we note that due to (\ref{rad17}) and smallness of $x$
\begin{equation}
\protect\label{fin10} 
t_{th}=\frac{4}{7}\sqrt{3\Psi(x)D}\,\tau_{th}^{7/4}=2\frac{\sqrt{2\Psi(x)}}{p}
\simeq 2\frac{\sqrt{\ln(2.35/x)}}{p x},
\end{equation}
for the soft part of spectrum. Here $x=h\nu/kT$, $\Psi(x)$ and $D$ are defined in
formulae \eqref{affat} and \eqref{rad16_1} correspondingly. 
Because $t_{th}$ is large for small values 
$x$ we can use approximation of the modified Bessel function of the second 
kind for large arguments
\[K_p(x)\approx\sqrt{\frac{\pi}{2x}}\,e^{-x},\]
and rewrite equation (\ref{rad12}) as follows
\[
J(\tau,x)=B_\nu \left(\frac{\tau}{\tau_{th}}\right)^2
\left[\frac{\Gamma(3/7)}{\Gamma(11/7)}z^{8/7}+\frac{4}{7\Gamma(11/7)}z^{1/14}e^{-2z}\right]=
\]
\begin{equation}
\protect\label{fin11}
=B_\nu \left(\frac{\tau}{\tau_{th}}\right)^2\left[2.32\,z^{8/7}+0.64\,z^{1/14}e^{-2z}\right],
\end{equation}
where 
\[
z=\frac{\sqrt{\ln (2.35/x)}}{p x}.
\]
Here we rewrite the dilution factor in terms of opacity using relation 
(\ref{rad13}). Clearly, the second term in the parenthesis of  formula (\ref{fin11}) is significant 
only where $z$ becomes small ($x$ becomes large) and the spectrum shape 
``adjusts'' to the black-body component. In turn, the first term of 
equation (\ref{fin11}) represents the power law component of the lower 
part of spectrum with the slope 6/7, which can be shown by simple similarity
 (see also T94)
\[B_\nu z^{8/7}\sim \frac{x^2}{x^{8/7}}=x^{6/7}.\]
Another important advantage of this term is that it vanishes for large values 
of $x$. This fact gives us opportunity to construct convenient and accurate 
formula for observational spectra fitting. We drop the second term in equation
 (\ref{fin11}) and adjust to the diluted black-body shape by means of quadratic 
power combination as follows
\begin{equation}
\protect\label{fin12}
J(\tau,x)=B_\nu \left(\frac{\tau}{\tau_{th}}\right)^2\left[1+5.34\,z^{16/7}\right]^{1/2}.
\end{equation}
Comparison of shapes given by formula (\ref{fin12}) with 
exact solution (\ref{rad12}) shows that they deviate from each other  by less 
than 2\% which is more than acceptable in  contemporary astrophysical observational 
data analysis. Using the explicit form of $z$ and the form of outgoing flux 
equation (\ref{fin12}) can be rewritten in the form 
\begin{equation}
\protect\label{fin13}
F_\nu=\frac{4\pi}{3}\frac{dJ_\nu}{d\tau}=\frac{8\pi}{3} 
B_\nu\frac{\tau_s}{\tau_{th}^2}\left\{1+5.34 
\left[\frac{\ln (2.35/x)}{p^2x^2}\right]^{8/7}\right\}^{1/2}.
\end{equation}
Equation (\ref{num1}) yields useful relationship for the dilution coefficient in 
the manner similar to (\ref{rad20})-(\ref{rad22})
\begin{equation}
\protect\label{fin14}
\frac{8\pi}{3} \frac{\tau_s}{\tau_{th}^2}=\frac{5.07\times10^{-5}r_{b,6}^{10/7}T_{b,9}^{10/17}p^{16/7}}{m^{11/7}v_{s,8}^{2/7}(2-Y_{He})^{1/7}}.
\end{equation}
Substituting results of parameter fitting (\ref{num2})-(\ref{num3}), we get 
\begin{equation}
\protect\label{fin15}
\frac{8\pi}{3} \frac{\tau_s}{\tau_{th}^2}= 
3.31\times10^{-3} r_{b,6}^{-0.36}\,T_{b,9}^{-0.29} m^{-0.17}(2-Y_{He})^{-0.58},
\end{equation}
\begin{equation}
\tag{\ref{num2}}
p=7.69\, T_{b,9}^{-0.84} r_{b,6}^{-0.89} m^{0.69} (2-Y_{He})^{-0.22}.
\end{equation}  
Here, again, $r_{b,6}$, $T_{b,9}$ and $m$ are the neutron star radius, surface temperature and mass in units of $10^6$ cm, $10^9$  Kelvin and solar mass respectively. Now we have provided our spectrum profile (\ref{fin13}) with expressions 
for parameter $p$ \eqref{num2} and the dilution factor \eqref{fin15}. These 
 three formulae constitute  the final analytical results of this paper.

\section{Discussion}

The model and derivations presented above assume that
plasma consists of fully ionized hydrogen and helium. In reality, this
assumption can be too simplistic. For instance, in the case of the recently 
discovered super-burst \citep{str}, a sufficient fraction of material should be
 represented by heavier elements.
These long and powerful bursts are also considered to be due to the nuclear
runaway burning in the carbon ``ocean'' under the neutron star surface. 
In this section we discuss how  our model can be adjusted for
study of this phenomenon. The approach as a whole does not change,
but some formulae have to be modified in order to account for the 
different plasma composition.

First, we note that for plasma which consists of a single ionized element, we
have for the mean molecular weight
\[
\mu=\frac{A}{1+Z},
\]
and for the electron number density 
\[
n_e=\frac{\rho}{A m_p}Z,
\]
where $A$ and $Z$ are the atomic weight and the atomic number of the corresponding element.
In the general case of heterogenous elements, each represented by weight abundance $Y_i$, we write
\begin{equation}
\protect\label{disc1}
\mu= \frac{1}{\sum Y_i(1+Z_i)/A_i}.
\end{equation}
\begin{equation}
\protect\label{disc2}
n_e=\frac{\rho}{m_p}\sum \frac{Z_i}{A_i}Y_i.
\end{equation}
In the hydrodynamic part of this study, these modifications will affect 
only the form of the terms 
and factors containing $Y_{He}$. In the radiation transfer section,
the form of $\alpha_{ff}/\alpha_T$ will require more careful treatment.
According to \citet{rbk79}, the free-free absorption coefficient is
\begin{equation}
\protect\label{disc4}
\alpha_{ff}=3.7\times 10^{8}\,T^{-1/2} \overline{Z^2n_i}n_e\,\nu^{-3}(1-e^{-h\nu/kT})\,\tilde{g}_{ff},
\end{equation}
where
\begin{equation}
\protect\label{disc5}
 \overline{Z^2n_i}=\sum Z_i^2n_i=\frac{\rho}{m_p} \sum \frac{Z_i^2}{A}Y_i.
\end{equation}
In the case of a hydrogen-helium plasma this factor is conveniently represented 
just by gas density, i. e.
\[ \overline{Z^2n_i}=n_H+4n_{He}=\frac{\rho}{m_p}(Y_H+Y_{He})=\frac{\rho}{m_p},\]
 which yields the equation \eqref{affat}. In general,
one should use  expressions \eqref{disc4} and \eqref{disc5} to find the correct
form of $\alpha_{ff}/\alpha_T$ relevant to the specific chemical composition.

 To be more instructive we conduct such a modification for the case when the plasma 
 has a substantial  carbon fraction. Using \eqref{disc1}, \eqref{disc2}
 and \eqref{disc5}, we write for hydrogen-helium-carbon gas
\begin{equation}
\protect\label{disc6}
\mu=\frac{12}{24Y_H+9Y_{He}+7Y_C}=\frac{4}{8-5Y_{He}-17Y_C/3},
\end{equation}
\begin{equation}
\protect\label{disc7}
n_e=\frac{\rho}{m_p}\left(1-\frac{Y_{He}+Y_C}{2}\right),
\end{equation}
and
\begin{equation}
\protect\label{disc8}
\overline{Z^2n_i}=\frac{\rho}{m_p}(Y_H+Y_{He}+3Y_C)=\frac{\rho}{m_p}(1+2Y_C).
\end{equation}
Correspondingly, in all formulae the factor 
$(2-Y_{He})$ will be replaced by $(2-Y_{He}-Y_C)$ and 
 $(8-5Y_{He})$ by $(8-5Y_{He}-17Y_C/3)$. Additionally, the right-hand side 
of equation \eqref{affat} has to be multiplied by the factor of $(1+2Y_C)$.
 Clearly, this modification will add the fifth free parameter $Y_C$ to
 the model. Using the general methodology outlined in this paper one 
should be able to produce solutions for the parameter $p$ and the 
dilution factor. The problem which
can arise from the inclusion of heavy elements is the
possibility for heavy ions to be only partly ionized. The ionization
degree can also vary throughout the atmosphere. Because full ionization
 and constancy of the gas's chemical composition are the basic 
assumptions of the adopted approach, we cannot explicitly include the 
effect of ionization in our model. Instead, it can be accounted for in a 
manner similar to our temperature profile correction.  First, the approximate atmospheric profiles can be
obtained by assuming full ionization.
Then, the ionization degree can be calculated by solving the Saha equation, and 
using this solution as  a zero-order approximation of the atmospheric 
temperature and the electron number density profiles.
Finally, one  should proceed by solving the hydrodynamic problem, 
in which the partial ionization of heavy elements is taken into account. 

For reasons mentioned
above, it is also a problem to include the proper physics for the transport
of heavy nuclei to the outer layers. Two major processes can contribute
to this element flow. Bulk motion mixing should dominate in the convection zone close to 
the bottom of the atmosphere. In the outer layers, a strong radiative push should govern the
process, because of the large resonance cross-sections of the heavy elements.
 The general problem of heavy ions mixing is quite difficult and 
requires a rigorous approach, which is out of scope of this paper.
 
As far as the boundary conditions are concerned, modeling of carbon 
nuclear flashes will require higher bottom temperatures.
Temperature of the carbon burning zone is argued to be about $10^{10}$ K
[see \citet{str}], which is close to the upper boundary for the bottom 
temperature $T_b$ used in our calculations. No peculiarities of the approach 
were detected in the case of very high bottom temperatures. 
Extremely high temperatures will require the correct form of the opacity 
coefficient $\kappa$ \citep{pac}, instead of equation \eqref{hydro2}, 
which represents a simplified formula for $\kappa$ in the case of modest
 temperatures.
 
Another important issue is the correct
accounting for the line emission of heavy elements which is detected
in the spectral analysis of super-bursts. \citet{str} argued that this
phenomena is due to reflection from the accretion disk during the
burst. One can estimate the disk heating time by using the standard Shakura-Sunyaev accretion disk model \citep{ss} 
and the fact that approximately 10\% of the burst luminosity is absorbed by
the inner part of the disk \citep{lapsun}. {\it Simple estimates give a timescale of less than a second assuming a mass-accretion rate of the order of
Eddington or less for the disk accretion regime, and a burst luminosity greater than 5\% of Eddington, which is
detected during several thousand seconds of observation of the super-burst in 4U 1820-30}
Consequently, the observed spectral feature of the K$\alpha$ line  should  rather be 
generated in the burst atmosphere than in the disk.  The disk  gains  the temperature of the X-ray radiation very quickly. 

Unfortunately, the origin and  behavior of the  spectral line features 
still remain unexplained. The authors plan to inlcude the spectral line effect 
in the relaxation method, in order to calculate the line emission during the X-ray burst
and to compare this with the observed spectra. 
  
Relativistic effects are usually negligible during the strong X-ray burst 
due to the significant radial expansion and the fact that the outgoing 
spectrum formation occurs at the outer layers of the atmosphere.
General relativistic effects become important  at the contraction stage
when the extended envelope  recedes close to the NS surface [see \citet{lpt}].
Haberl \& Titarchuk (1995) applied the full general relativity approach 
for a derivation of NS mass-radius relation in 4U 1820-30 using EXOSAT observations 
and the T94 model.

\section{Conclusion}
This paper follows a common idea of the last decade to fit observational 
and numerical spectra with some model, mostly black-body shapes, to obtain
 spectral softening/hardening factors \citep{lth}. We improve this 
technique in several ways. We use for fitting more realistic 
non-blackbody spectral profile, which accounts for the observed 
power law soft excess of X-ray burster spectra. The temperature profile
is corrected by solving the temperature equation. The 
existence of the isothermal photosphere during X-ray bursts is confirmed
numerically and analytically. 
Finally, we analytically obtain the multiplicative (dilution) factor which 
is not a parameter of fitting anymore but self-consistently incorporated 
in the model. 

We show how the theoretical study of radiatively driven wind phenomenon can 
produce useful techniques for analyzing observational data. It can fulfill
the needs of new emerging  branches of observational X-ray astronomy 
 such as a very promising 
discovery of super-bursts \citep{str}, which exhibit photospheric 
expansion and spectral modifications relevant to extended atmospheres.
We present the analytical theory of strong X-ray bursts, 
which include effects of Comptonization and free-free absorption. 
Partly presented in some earlier publications, this area of the study of the X-ray burst spectral 
formation was lacking a detailed and self-consistent account.
 We use numerical simulation to validate our analytical theory 
and to link our solution to energy axes.  
We show how this  information can be extracted from spectral data.
 We provide the analytical  expression for the X-ray burst spectral shape, 
 which depends only upon input physical 
parameters of the problem: neutron star mass, radius, surface temperature and elemental abundance. 
Expressions for color ratios and dilution coefficient are also given.

Authors thank Peter Becker for valuable comments and 
suggestions which improved the paper. 
We are greatful to Menas Kafatos for encouragement and to 
Center for Earth Science and Space Research (GMU) for the support of 
this research. We appreciate the thorough analysis of the presented work by 
the referee.

\appendix

\section{Analytic solution for the radiative transfer problem} 

We look for the solution of equation (\ref{rad8}) in the form
\begin{equation}
\protect\label{A1}
J(\tau,x)=\left(\frac{\tau}{\tau_{th}}\right)^2B_\nu(\tau_{th})+\tilde{J}(\tau,x).
\end{equation}
The basic idea is to separate the high-frequency (diluted black body) and 
the low-frequency $\tilde{J}(\tau,x)$ parts of spectrum, where different 
physical processes dominate. Kompaneets operator $L_\nu$ acting upon black 
body shape vanishes and we neglect $L_\nu(\tilde{J})$, which allows 
us to get the solution of radiative transfer problem analytically. At 
this point $\tau_{th}$ is a parameter of the problem. The algorithm of 
determination of $\tau_{th}$ will be described separately. Substituting (\ref{A1}) into (\ref{rad8}) we found for $\tilde{J}(\tau,x)$ 
\begin{equation}
\frac{\partial}{\partial \tau}\left(\frac{1}{\tau}\frac{\partial \tilde{J}}{\partial \tau}\right)-\frac{3}{\tau}\frac{\alpha_{ff}}{\alpha_T}\tilde{J}=-\frac{3}{\tau}\frac{\alpha_{ff}}{\alpha_T}B_\nu\left[1-\left(\frac{\tau}{\tau_{th}}\right)^2\right],
\end{equation}
with a boundary condition
\begin{equation}
\tilde{J}_\nu|_{\tau=\tau_{th}}=0. 
\end{equation}
The solution satisfying this condition is presented by
\begin{equation}
\protect\label{A2}
\tilde{J}(\tau,x)=\frac{1}{pW}y_1(\tau)\int_0^{\tau_{th}}y_2(\tau)f(\tau)\,d\tau,
\end{equation}
where $ p(\tau)=\frac{1}{\tau}$ and $W(\tau)$ is the Wronskian
\begin{equation}
W=\left|\begin{array}{cc}
y_1&y_2\\
y_1'&y_2'
\end{array}\right|=-\frac{7}{4}\tau.
\end{equation}
Thus the product
\[
pW=-\frac{7}{4}.
\]
Functions $y_1(x)$ and $y_2(x)$ are 
\begin{equation}
y_1(\tau)=\tau I_{4/7}\left(\frac{4}{7}\sqrt{3D\Psi(x)}\,\tau^{7/4}\right),
\end{equation}
\begin{equation}
y_2(\tau)=\tau K_{4/7}\left(\frac{4}{7}\sqrt{3D\Psi(x)}\,\tau^{7/4}\right),
\end{equation}
where $I_\nu(x)$ and $K_\nu(x)$ are modified Bessel function of the first and 
the second types respectively.

The function $f(\tau)$ in (\ref{A2}) is the right hand side 
of equation (A2), namely
\begin{equation}
f(\tau)=-\frac{3}{\tau}\frac{\alpha_{ff}}{\alpha_T}=-3\,\tau^{1/2}D\Psi(x)B_\nu\left[1-\left(\frac{\tau}{\tau_{th}}\right)^2\right].
\end{equation}
We introduce a new variable $t$
\begin{equation}
t=\frac{4}{7}\sqrt{3D\Psi(x)}\,\tau^{7/4},
\end{equation}
and rewrite  solution (\ref{A2}) as
\begin{equation}
\protect\label{A3}
\tilde{J}(t,x)=B_\nu t^{4/7}I_{4/7}(t)\int_0^{t_{th}}t^{3/7}K_{4/7}(t)\left[1-\left(\frac{t}{t_{th}}\right)^{8/7}\right]dt.
\end{equation}
Using the properties of modified Bessel functions
\[\int x^p K_{p-1} dx=-x^p K_{p}+C\quad\mathrm{and}\quad K_p=K_{-p}\]
we evaluate the integrals in (\ref{A3})
\[ \int_0^{t_{th}}t^{3/7}K_{4/7}(t)\,dt=-t^{3/7}K_{3/7}(t)\Bigl|_0^{t_{th}}=\frac{\Gamma\left(\frac{3}{7}\right)}{2^{4/7}}-t_{th}^{3/7}K_{3/7}(t_{th}), \]
\[ \int_0^{t_{th}}t^{11/7}K_{4/7}(t)\,dt=-t^{11/7}K_{11/7}(t)\Bigl|_0^{t_{th}}=\Gamma\left(\frac{11}{7}\right)2^{4/7}-t_{th}^{11/7}K_{11/7}(t_{th}). \]
Finally $\tilde{J}(t,x)$ takes the form
\[
\tilde{J}(t,x)=B_\nu t^{4/7}I_{4/7}(t)\left[\frac{\Gamma\left(\frac{3}{7}\right)}{2^{4/7}}- \frac{\Gamma\left(\frac{11}{7}\right)2^{4/7}}{t_{th}^{8/7}}+t_{th}^{3/7}(K_{11/7}(t_{th})-K_{3/7}(t_{th}))\right]=
\]
\begin{equation}
\tilde{J}(t,x)=B_\nu t^{4/7}I_{4/7}(t)\left[\frac{\Gamma\left(\frac{3}{7}\right)}{2^{4/7}}- \frac{\Gamma\left(\frac{11}{7}\right)2^{4/7}}{t_{th}^{8/7}}+\frac{8}{7}\,t_{th}^{-4/7}K_{4/7}(t_{th})\right].
\end{equation}
The last formula is a solution of equation (\ref{rad8}). 
We can simplify this form by noting that we are interested in the solution 
in the outer layers of atmosphere (emergent spectrum) where $\tau \rightarrow 0$ and 
$t \rightarrow 0$, and we can use the asymptotic form for small arguments  
\[I_p(x)\approx\frac{1}{\Gamma(p+1)}\left(\frac{x}{2}\right)^p.\]
Making this substitution and putting the result into expression for $J(\tau,x)$ we 
 find that second term in $\tilde{J}(\tau,x)$ cancels with diluted blackbody term in 
  $J(\tau,x)$, which takes the form
\begin{equation}
J(t,x)= B_\nu \frac{t^{8/7}}{2^{4/7}\Gamma\left(\frac{11}{7}\right)}\left[\frac{\Gamma\left(\frac{3}{7}\right)}{2^{4/7}}+\frac{8}{7}\,t_{th}^{-4/7}K_{4/7}(t_{th})\right].
\end{equation}
\bigskip

\section{Solution of the temperature equation}

Substituting relation (\ref{rad14}) into the equation of radiative diffusion, 
multiplying it by $x^2$ and integrating over the energy range 
from $x_*$ to $\infty$ we get
\begin{equation}
\protect\label{e1}
\frac{1}{3}\left(\frac{d^2R}{d\tau^2}-\frac{1}{\tau}\frac{dR}{d\tau}\right)
\int_{x_*}^\infty\frac{x^2}{e^x-1}\,dx=[R(\tau)-1]\int_{x_*}^\infty\frac{x^2}{e^x-1}
\frac{\alpha_{ff}}{\alpha_T}\,dx,
\end{equation}
where integrals can be approximated as
\[\int_{x_*}^\infty\frac{x^2}{e^x-1}\approx \int_0^\infty x^2e^{-x}dx=2\]
and, noting that $\alpha_{ff}/\alpha_T=D\Psi(x)\,\tau^{3/2}\approx 
D\,\tau^{3/2}\tilde g(x)/x^2$ we obtain
\[ \int_{x_*}^\infty\frac{x^2}{e^x-1}\frac{\alpha_{ff}}{\alpha_T}\,dx\approx D\,\tau^{3/2}\int_{x_*}^\infty\frac{\tilde{g}(x)}{x}\,dx\approx \frac{1}{4}\ln^2\frac{2.25}{x_*}D\,\tau^{3/2}.\]
Here we used the fact that $\tilde{g}(x)\approx \frac{1}{2}\ln(2.35/x)$. 
The equation for $R(\tau)$ gets the form
\begin{equation}
\frac{d^2R}{d\tau^2}-\frac{1}{\tau}\frac{dR}{d\tau}=\frac{3}{8}\ln^2\frac{2.25}{x_*}D\tau^{3/2}[R(\tau)-1]=\tilde{D}\tau^{3/2}[R(\tau)-1].
\end{equation}
Boundary conditions for this equation are
\begin{equation}
\begin{array}{cc}
\tau\rightarrow 0&R(\tau)\rightarrow 0,\\
\tau\rightarrow \infty&R(\tau)\rightarrow 1.
\end{array}
\end{equation}
A general solution of equation (29) is
\begin{equation}
R(\tau)=1+\tau Z_{4/7}\left(\frac{4}{7}i\sqrt{\tilde{D}}\tau^{7/4}\right),
\end{equation}
where $Z_{4/7}(z)=c_1K_{4/7}(z)+c_2I_{4/7}(z)$. In derivation of this formula we take
into account a well known theorem from ODE theory that general solution of inhomogeneous 
ODE is the sum of a general solution of the corresponding homogeneous ODE and some particular 
solution of the inhomogeneous ODE, which is chosen equals to unity in our case. 
The second boundary condition and the fact that
\begin{displaymath}
\begin{array}{cc}
K_{4/7}(z)\rightarrow 0&z\rightarrow \infty,\\
I_{4/7}(z)\rightarrow \infty&z\rightarrow \infty
\end{array}
\end{displaymath}
leave only $c_1$ nonzero, and the first boundary condition gives us the value for $c_1$, namely
\[c_1=-\frac{1}{ _\tau\varinjlim_{0}\tau K_{4/7}
\left(\frac{4}{7}\sqrt{\tilde{D}}\,\tau^{7/4}\right)}=
-\frac{2^{3/7}}{\Gamma\left(\frac{4}{7}\right)\tau_{th}},
\]
where we put $\tau_{th}=\left(\frac{4}{7}\sqrt{\tilde{D}}\right)^{-4/7}$. 
Then $R(\tau)$ reduces to
\begin{equation}
R(\tau)=1-\frac{2^{3/7}}{\Gamma\left(\frac{4}{7}\right)}\frac{\tau}{\tau_{th}} 
K_{4/7}\left[\left(\frac{\tau}{\tau_{th}}\right)^{7/4}\right].
\end{equation}
Taylor series expansion of $K_{4/7}$ over $\tau/\tau_{th}$  yields for $R(\tau)$ useful relation
\begin{equation}
\protect\label{B6}
R(\tau)=\frac{\Gamma\left(\frac{3}{7}\right)}{2^{8/7}
\Gamma\left(\frac{11}{7}\right)}\left(\frac{\tau}{\tau_{th}}\right)^2.
\end{equation}

\section{Condition at the sonic point (derivation of $Y_s$)}

We can rewrite the partial derivative in (\ref{hydro21}) using the obvious relation
\[
\left(\frac{\partial P}{\partial \rho}\right)_\Xi=
\frac{\partial P}{\partial T}\left(\frac{\partial T}{\partial \rho}\right)_\Xi+\frac{\partial P}{\partial y}\left(\frac{\partial y}{\partial \rho}\right)_\Xi.
\]

Differentiating  the equation of state (\ref{hydro9}) we obtain derivatives of pressure
\[ \frac{\partial P}{\partial T}=\left(1+\frac{1}{y}\right)\frac{4aT^3}{3}
\quad\mathrm{and}\quad\frac{\partial P}{\partial y}=-\frac{1}{y^2}\frac{aT^4}{3},\]
and differentiating (\ref{hydro24}) with respect to $\rho$ we get
\[\left(\frac{\partial T}{\partial \rho}\right)_\Xi=
-\frac{T}{\lambda}\left(\frac{1}{y}+4\right)
\left(\frac{\partial y}{\partial \rho}\right)_\Xi.
\]
From the other hand differentiation of (\ref{hydro10}) gives us
\[\frac{a\mu m_p}{3k}\left(\frac{3T^2}{y}
\frac{\partial T}{\partial \rho}-\frac{T^3}{y^2}\frac{\partial y}{\partial \rho}\right)=1.\]
Combination with the previous equation it yields:
\[\left(\frac{\partial y}{\partial \rho}\right)_\Xi=
-\frac{3k}{a\mu m_p}\frac{y^2}{T^3}\left(\frac{\lambda}{\lambda+3(1+4y)}\right),\]
\[\left(\frac{\partial T}{\partial \rho}\right)_\Xi=
\frac{3k}{a\mu m_pT^2}\left(\frac{y(1+4y)}{\lambda+3(1+4y)}\right).\]

Now, combining all found derivatives we have
\[
\left(\frac{\partial P}{\partial \rho}\right)_\Xi=
\frac{k}{\mu m_p}\left[4\left(1+\frac{1}{y}\right)\left(\frac{y(1+4y)}
{\lambda+3(1+4y)}\right)+\frac{\lambda}{\lambda+3(1+4y)}\right]T\]
or in a more compact form
\begin{equation}
\label{appc1}
\left(\frac{\partial P}{\partial \rho}\right)_\Xi=\frac{k}{\mu m_p}
\left[\frac{\lambda+4(1+y)(1+4y)}{\lambda+3(1+4y)}\right]T,
\end{equation}
which is, in fact, a sonic point condition (\ref{hydro21}).

\section{Reduction  of Hydrodynamical Problem to First-Order ODE}

We derive an expression for the derivative of velocity $v_y$ through  $v$ and $y$. 

We substitute the temperature  profile found in (\ref{hydro24}) to (\ref{hydro6}) 
  to obtain $\rho$ as a function of $y$
\[ \rho=\rho(y)=\frac{a\mu m_p T_b^3}{3 k}\,y^{-3/\lambda-1}
\exp \left[-\frac{12(y-1)}{\lambda}\right].
\]
Using this expression for $\rho(y)$ and equation (\ref{hydro13}) we get
\[
r=r(v,y)=\left(\frac{\Phi}{4\pi}\right)^{1/2}\rho^{-\frac{1}{2}}v^{-\frac{1}{2}}=
\left(\frac{3\Phi k}{4\pi a \mu m_p T_b^3}\right)^{1/2}
y^{\frac{3}{2\lambda}+\frac{1}{2}}\exp \left[\frac{6(y-1)}{\lambda}\right]v^{-\frac{1}{2}}.
\]
Then we get derivatives
\[ \frac{dr}{dy}=r\left[\left(\frac{3}{2\lambda}+
\frac{1}{2}\right)\frac{1}{y}+\frac{6}{\lambda}\right],
\]
\[
 \frac{dr}{dv}=-\frac{r}{2v}.
\]
Differentiation of (\ref{hydro24}) also gives us 
\[
\frac{dT}{dy}=T_{b}\,y^{-1/\lambda}\exp \left[-\frac{4(y-1)}{\lambda}\right]\left(-\frac{1}{\lambda y}-\frac{4}{\lambda}\right)=-\frac{T}{\lambda}\left(4+\frac{1}{y}\right).
\]
By a combination of all these derivatives we obtain
\[
\frac{dT}{dr}=\frac{dT}{dy}\frac{dy}{dr}=
\frac{dT}{dy}\left(\frac{\partial r}{\partial y}+
\frac{\partial r}{\partial v}\frac{d v}{d y}\right)^{-1}=
-\frac{2T(4y+1)}{r\lambda y}\left[\left(\frac{3}{\lambda}+1
\right)\frac{1}{y}+\frac{12}{\lambda}-\frac{v'}{v}\right]^{-1}.
\]
Substitution of it into (\ref{hydro6}) yields
\[
 L_{r}=-\frac{16\pi c k r^{2} y}{\mu m_p \kappa}\frac{dT}{dr}
 =\frac{32\pi c k }{\mu m_p \lambda \kappa_0 (2-Y_{He})}
 \frac{(4y+1)(1+\alpha T)Tr}{\left[\left(3/\lambda+1\right)
 1/y+12/\lambda-v'/v\right]}.\]
But from (\ref{hydro14}) we also have
\[
 L_{r}=\Psi-\Phi \left(h+\frac{v^2}{2}-\frac{G M_{ns}}{r}\right).\]
Equating the last two expressions for $L_r$ we finally find $v_y$ as
\[v_y^{\prime}=f(v,y)=v\left[\left(1+3\frac{1+4y}{\lambda}\right)\frac{1}{y}-
75.2\frac{rT(8-5Y_{He})(1+4y)(1+\alpha T)}{\lambda r_{b,6}T_{b,9} 
(\Psi/\Phi-v^2/2-h+G M_{ns}/r)}\right].\]
Here $r_{b,6}$ and $T_{b,9}$ represent neutron star radius and the bottom temperature of atmosphere in terms of $10^6$ cm and $10^9$ K correspondingly.

\begin{deluxetable}{lrrrr@{.}lrr@{.}lcccc}
\tabletypesize{\scriptsize}
\tablecaption{Parameters for model $m=1.5$, $R_{ns}=13.5$ km, $Y_{He}=1$ \label{tbl-1}}
\tablewidth{0pt}
\tablehead{
\colhead{$T_b$} & \colhead{$kT$}   & \colhead{$T/T_{0}$}   &
\colhead{$\tau_{th}$} &
\multicolumn{2}{c}{$p$}  & \colhead{$\Phi$ \tablenotemark{a}} & \multicolumn{2}{c}{$M_{env}$} &
\colhead{$T_{s}$}     & \colhead{$v_s$}  &
\colhead{$r_{s}$}   & \colhead{$r_{ph}$}  \\
\colhead{$10^9$K} & \colhead{keV}   & \colhead{}   &
\colhead{$$} &
\multicolumn{2}{c}{}  & \colhead{ } &  \multicolumn{2}{c}{$10^{22}$g} &
\colhead{$0.1$ keV}     & \colhead{$10^3$ km/s }  &
\colhead{$10^3$km}   & \colhead{$10^3$km}  

}
\startdata
7.0 & 0.197 &  0.099 &  45.1  & 1&56 & 93.9  & 173&3 &  0.21  & 1.31  & 57.9  & 4.16 \\ 
6.5 & 0.208 &  0.105 &  44.7  & 1&64 & 87.2  & 128&9 &  0.22  & 1.39  & 51.6  & 3.71 \\ 
6.0 & 0.221 &  0.111 &  44.1  & 1&75 & 80.5  & 93&6 &  0.24  & 1.48  & 45.6  & 3.28 \\ 
5.5 & 0.236 &  0.118 &  43.3  & 1&87 & 73.8  & 66&1 &  0.25  & 1.58  & 39.9  & 2.88 \\ 
5.0 & 0.252 &  0.127 &  42.4  & 2&02 & 67.1  & 45&1 &  0.27  & 1.70  & 34.5  & 2.50 \\ 
4.5 & 0.272 &  0.137 &  41.5  & 2&20 & 60.4  & 29&6 &  0.29  & 1.84  & 29.4  & 2.14 \\ 
4.0 & 0.295 &  0.148 &  40.3  & 2&42 & 53.7  & 18&5 &  0.32  & 2.01  & 24.6  & 1.80 \\ 
3.5 & 0.324 &  0.163 &  39.0  & 2&71 & 46.9  & 10&9 &  0.35  & 2.22  & 20.1  & 1.48 \\ 
3.0 & 0.359 &  0.180 &  37.4  & 3&09 & 40.2  & 5&86 &  0.39  & 2.50  & 16.0  & 1.19 \\ 
2.5 & 0.406 &  0.204 &  35.6  & 3&60 & 33.5  & 2&83 &  0.44  & 2.85  & 12.2  & 0.92 \\ 
2.0 & 0.470 &  0.236 &  33.4  & 4&34 & 26.8  & 1&16 &  0.51  & 3.36  & 8.83  & 0.67 \\ 
1.75 & 0.510 &  0.257 &  32.2  & 4&85 & 23.5  & 0&68 &  0.55  & 3.69  & 7.29  & 0.56 \\ 
1.5 & 0.565 &  0.284 &  30.8  & 5&53 & 20.1  & 0&37 &  0.61  & 4.12  & 5.86  & 0.45 \\ 
1.25 & 0.634 &  0.318 &  29.2  & 6&44 & 16.8  & 0&179 &  0.68  & 4.69  & 4.53  & 0.36 \\ 
1.1 & 0.686 &  0.344 &  28.0  & 7&18 & 14.8  & 0&108 &  0.74  & 5.12  & 3.80  & 0.30 \\ 
1.0 & 0.727 &  0.365 &  27.2  & 7&78 & 13.4  & 0&074 &  0.78  & 5.47  & 3.33  & 0.27 \\ 
0.9 & 0.775 &  0.389 &  26.3  & 8&51 & 12.1  & 0&049 &  0.84  & 5.87  & 2.89  & 0.23 \\ 
0.8 & 0.831 &  0.417 &  25.3  & 9&40 & 10.7  & 0&031 &  0.90  & 6.36  & 2.46  & 0.20 \\ 
0.7 & 0.898 &  0.451 &  24.3  & 10&5 & 9.4  & 0&018 &  0.97  & 6.95  & 2.06  & 0.17 \\ 
0.6 & 0.981 &  0.492 &  23.0  & 12&0 & 8.0  & 0&010 &  1.06  & 7.69  & 1.68  & 0.14 \\ 
0.5 & 1.086 &  0.545 &  21.6  & 14&0 & 6.7  & 0&005 &  1.17  & 8.65  & 1.33  & 0.11 \\ 
0.4 & 1.224 &  0.614 &  19.9  & 17&0 & 5.4  & 0&002 &  1.32  & 9.95  & 1.01  & 0.09 \\ 
0.3 & 1.417 &  0.711 &  17.8  & 21&9 & 4.0  & 0&001 &  1.53  & 11.8  & 0.71  & 0.07 \\ 
\enddata

\tablenotetext{a}{ $\Phi $ is  in units of the critical mass-loss rate, i. e. divided by  $L_{E}/c^2$}


\end{deluxetable}

\clearpage

\begin{deluxetable}{lrrrr@{.}lr@{.}lr@{.}lcr@{.}lr@{.}lr}
\tabletypesize{\scriptsize}
\tablecaption{Parameters for model $m=2.6$, $R_{ns}=14.0$ km, $Y_{He}=1$. \label{tbl-2}}
\tablewidth{0pt}
\tablehead{
\colhead{$T_b$} & \colhead{$kT$}   & \colhead{$T/T_{0}$}   &
\colhead{$\tau_{th}$} &
\multicolumn{2}{c}{$p$}  & \multicolumn{2}{c}{$\Phi$} & \multicolumn{2}{c}{$M_{env}$} &
\colhead{$T_{s}$}     & \multicolumn{2}{c}{$v_s$}  &
\multicolumn{2}{c}{$r_{s}$}   & \colhead{$r_{ph}$}  \\
\colhead{$10^9$K} & \colhead{keV}   & \colhead{}   &
\colhead{$$} &
\multicolumn{2}{c}{}  & \multicolumn{2}{c}{} & \multicolumn{2}{c}{$10^{22}$g} &
\colhead{$0.1$ keV}     & \multicolumn{2}{c}{$10^3$ km/s }  &
\multicolumn{2}{c}{$10^3$km}   & \colhead{$10^3$km}  

}
\startdata
7.0 & 0.248 &  0.110 &  45.1  & 2&12 & 56&2  & 115&8 &  0.242  & 1&80  & 53&4  & 3.53 \\ 
6.5 & 0.261 &  0.116 &  44.5  & 2&24 & 52&2  & 86&1 &  0.256  & 1&90  & 47&7  & 3.15 \\ 
6.0 & 0.277 &  0.123 &  43.6  & 2&40 & 48&2  & 62&6 &  0.271  & 2&02  & 42&2  & 2.80 \\ 
5.5 & 0.294 &  0.131 &  42.6  & 2&58 & 44&1  & 44&2 &  0.288  & 2&16  & 36&9  & 2.46 \\ 
5.0 & 0.313 &  0.139 &  41.5  & 2&80 & 40&1  & 30&2 &  0.308  & 2&32  & 31&9  & 2.15 \\ 
4.5 & 0.337 &  0.150 &  40.3  & 3&07 & 36&1  & 19&8 &  0.331  & 2&52  & 27&2  & 1.84 \\ 
4.0 & 0.364 &  0.162 &  39.0  & 3&40 & 32&1  & 12&4 &  0.359  & 2&75  & 22&8  & 1.56 \\ 
3.5 & 0.398 &  0.177 &  37.4  & 3&82 & 28&1  & 7&27 &  0.394  & 3&04  & 18&7  & 1.29 \\ 
3.0 & 0.440 &  0.196 &  35.7  & 4&36 & 24&1  & 3&93 &  0.437  & 3&41  & 14&9  & 1.04 \\ 
2.5 & 0.495 &  0.220 &  33.8  & 5&12 & 20&1  & 1&90 &  0.494  & 3&89  & 11&4  & 0.81 \\ 
2.0 & 0.570 &  0.254 &  31.5  & 6&21 & 16&1  & 0&78 &  0.572  & 4&58  & 8&24  & 0.59 \\ 
1.75& 0.620 &  0.276 &  30.2  & 6&97 & 14&0  & 0&46 &  0.623  & 5&03  & 6&80  & 0.50 \\ 
1.5 & 0.682 &  0.304 &  28.7  & 7&97 & 12&0  & 0&25 &  0.687  & 5&61  & 5&47  & 0.40 \\ 
1.25& 0.763 &  0.339 &  27.0  & 9&33 & 10&0  & 0&121 & 0.770  & 6&38  & 4&24  & 0.32 \\ 
1.1 & 0.824 &  0.366 &  25.9  & 10&4 & 8&83  & 0&073 &  0.833  & 6&96  & 3&56  & 0.27 \\ 
1.0 & 0.872 &  0.388 &  25.1  & 11&3 & 8&03  & 0&050 &  0.883  & 7&43  & 3&12  & 0.24 \\ 
0.9 & 0.928 &  0.413 &  24.2  & 12&4 & 7&22  & 0&033 &  0.940  & 7&98  & 2&71  & 0.21 \\ 
0.8 & 0.993 &  0.442 &  23.3  & 13&7 & 6&42  & 0&021 &  1.008  & 8&63  & 2&31  & 0.18 \\ 
0.7 & 1.072 &  0.477 &  22.2  & 15&4 & 5&62  & 0&012 &  1.089  & 9&43  & 1&94  & 0.15 \\ 
0.6 & 1.169 &  0.520 &  21.0  & 17&6 & 4&82  & 0&007 &  1.188  & 10&4  & 1&59  & 0.13 \\ 
0.5 & 1.292 &  0.575 &  19.7  & 20&6 & 4&01  & 0&003 &  1.313  & 11&7  & 1&26  & 0.10 \\ 
0.4 & 1.455 &  0.647 &  18.1  & 24&9 & 3&21  & 0&001 &  1.479  & 13&5  & 0&95  & 0.08 \\ 
0.3 & 1.682 &  0.749 &  16.2  & 32&0 & 2&41  & 0&0005 &  1.710  & 16&0  & 0&67  & 0.06 \\ 
\enddata



\end{deluxetable}

\end{document}